\documentclass[aps,prd,showpacs,superscriptaddress,nofootinbib,floatfix]{revtex4}
\usepackage{amsmath}
\usepackage{amssymb,amsfonts}

\DeclareFontFamily{OT1}{rsfs}{}
\DeclareFontShape{OT1}{rsfs}{m}{n}{<-7> rsfs5 <7-10> rsfs7 <10->rsfs10}{}
\DeclareMathAlphabet{\mycal}{OT1}{rsfs}{m}{n}
\newcommand{\scri}{\mycal I}

\usepackage{graphicx}
\usepackage{color}
\usepackage{epsfig}

\def\be{\begin{equation}}
\def\ee{\end{equation}}
\def\ba{\begin{eqnarray}}
\def\ea{\end{eqnarray}}

\newcommand{\Ham}{H}
\newcommand{\tpb}{\widetilde{p}_b}
\newcommand{\tpc}{\widetilde{p}_c}
\newcommand{\tb}{\widetilde{b}}
\newcommand{\tc}{\widetilde{c}}
\newcommand{\lp}{l_{\text{Pl}}}
\newcommand{\mplanck}{m_{\text{Pl}}}

\begin{document}

\title{Loop Quantum Dynamics of the Schwarzschild Interior}

\author{Christian G.~B\"ohmer}
\email{c.boehmer@ucl.ac.uk}
\affiliation{Department of Mathematics, University College London,
             Gower Street, London, WC1E 6BT, UK}
\affiliation{Institute of Cosmology \& Gravitation,
             University of Portsmouth, Portsmouth PO1 2EG, UK}
\author{Kevin Vandersloot}
\email{kevin.vandersloot@port.ac.uk}
\affiliation{Institute of Cosmology \& Gravitation,
             University of Portsmouth, Portsmouth PO1 2EG, UK}
\affiliation{Institute for Gravitational Physics and Geometry,
             Physics Department, Pennsylvania State University,
	     University Park, PA 16802, U.S.A.}

\date{\today}

\begin{abstract}
We examine the Schwarzschild interior of a black hole, incorporating
quantum gravitational modifications due to loop quantum gravity. We
consider an improved loop quantization using techniques that have
proven successful in loop quantum cosmology. The central Schwarzschild
singularity is resolved and the implications for the fate  of an in-falling
test particle in the interior region is discussed. The singularity is
replaced by a Nariai type Universe. We discuss the resulting conformal
diagram, providing  a clear geometrical interpretation
of the quantum effects.
\end{abstract}


\maketitle

\section{Introduction}
A long held expectation  is that
the singularities predicted from  general relativity signal
a breakdown of the classical theory requiring a more proper
accounting of the quantum effects of gravity. The situation is analogous
to
 Hydrogen atom in ordinary  mechanics, where classically the
electron is expected to in-spiral towards the proton leading to a singularity.
This behavior is cured when quantum mechanics is taken into account
leading to a stable non-singular Hydrogen atom. In general relativity, two
of the most relevant forms of singularities are the big-bang cosmological
singularity, as well as black-hole singularities. A central question to be
answered by any quantum theory of gravity is whether these singularities are
regulated by incorporating  quantum effects. Furthermore, if a quantum
theory of gravity were to resolve the classical singularities, it would be
of paramount interest as to what replaces them dynamically.

A leading candidate theory of quantum gravity is known as loop quantum
gravity (for reviews see~\cite{ash10,book,Thiemann:2001yy}). The application
of loop quantum gravity techniques to cosmological models is known as
loop quantum cosmology~\cite{Bojowald:2006da} which has led to a resolution
of the big-bang singularity, replacing it with a big-bounce for homogeneous
and isotropic models~\cite{Ashtekar:BBII,Newkplus1,Vandersloot:2006ws}. These
results provide tantalizing first hints that loop quantum gravity
may indeed provide the singularity resolution hoped for in a quantum
theory of gravity. Thus the next step is to consider loop quantum gravity
for the black hole scenario.

The simplest first step in considering loop quantum black holes consists of
examining the interior of a Schwarzschild black hole. There, the temporal
and radial coordinates flip roles, and the interior becomes a Kantowski-Sachs cosmological
model whereby the metric components are homogeneous and only time dependent. The interior
therefore can be quantized in a similar fashion as loop quantum cosmology
leading to the possibility that  the singularity is resolved
as in the cosmological case. The loop quantization of the Schwarzschild
interior has been initially developed in~\cite{Ashtekar:2005qt} and \cite{Modesto:2004wm}. There,
the quantization indicates that the quantum Einstein equations are non-singular
in a similar way loop quantum cosmology was originally shown to be
non-singular~\cite{Bojowald:2001xe}. However, the question of what replaces
the black hole singularity is not answered.

Therefore, we attempt to provide an answer as to what dynamics are predicted
from the loop quantization of the Schwarzschild interior. Initial attempts in
this direction are given in~\cite{Modesto:2006mx} which is based
on the quantum Einstein equations of the original quantization of~\cite{Ashtekar:2005qt}.
However, this quantization is the direct analog of the original formulations
of loop quantum cosmology which have been shown to not have a good semi-classical
limit~\cite{Ashtekar:BBII}. In particular, a crucial parameter in the construction
of the Hamiltonian constraint operator is taken to be a constant in the early
constructions and it is this assumption that can spoil the semi-classical limit
in loop quantum cosmology. An improved quantization of loop quantum cosmology whereby
the parameter is assumed to explicitly depend on the scale factor
has been applied in~\cite{Ashtekar:BBII} and leads to the correct
semi-classical behavior as well as more physically sensible quantum corrections
to the standard cosmological dynamics.

In this paper, we  consider the improved quantization scheme
of loop quantum cosmology applied to the loop quantization of the Schwarzschild
interior. We will consider the resulting modifications to the classical interior
by examining effective semi-classical  equations that incorporate quantum
effects. While we do not have a complete handling of the physical sector of
the quantum theory owing to its complexity, the effective semi-classical description
is intended to be an approximate shortcut to the physical predictions of
the quantum theory. In the framework of the effective description, we will
show that the singularity is resolved and discuss the implications for
an in-falling test particle as well as discuss the resulting conformal diagram for the non-singular
interior region.

\section{Classical Framework}

\subsection{Classical Hamiltonian}

The interior of the Schwarzschild geometry can be
described by a Kantowski-Sachs homogeneous model whereby
the line element is given by
\be\label{metric}
	ds^2 = -N(t)^2 dt^2 + g_{xx}(t)\, dx^2 + g_{\Omega \Omega}(t)\, d\Omega^2
\ee
with $N(t)$ being the freely specifiable lapse function and
 $d\Omega^2$ representing the unit two-sphere metric given
in polar coordinates
\be
	d\Omega^2 = d\theta^2 + \sin^2\theta \,d\phi^2 \,.
\ee
The topology of the spatial slices is ${\mathbb R} \times S^2$ hence the  coordinates
range from $x\in {\mathbb R},\, \theta \in[0, \pi],\, \phi \in [0, 2\pi]$.
In the standard Schwarzschild solution, the metric components are given by
\be \label{Swcl}
	N(t)^2 = \bigg(\frac{2m}{t}-1\bigg)^{-1}, \quad g_{xx}(t) = \bigg(\frac{2m}{t}-1\bigg), \quad g_{\Omega \Omega}(t) = t^2
\ee
for $t$ in the range $t \in [0, 2m]$, where $m$ is the mass of the black hole.

Loop quantum gravity is based on a connection-triad Hamiltonian
formulation of general relativity. Considering connections and triads
that are symmetric under the Kantowski-Sachs symmetry group
$\mathbb{R} \times \text{SO}(3)$, the connection $A_a^i$ and
triad  $E^a_i$  after gauge fixing of the Gauss constraint are
of the form~\cite{Ashtekar:2005qt}
\ba
	A_a^i  \tau_i dx^a &=& \tilde{c} \tau_3 dx + \tilde{b} \tau_2 d\theta
	- \tilde{b} \tau_1 \sin\theta d\phi + \tau_3 \cos\theta d\phi \\
	E^a_i \tau_i \partial_a &=& \tilde{p}_c \tau_3 \sin\theta \,\partial_x
	+\tilde{p}_b \tau_2 \sin\theta \, \partial_{\theta} - \tilde{p}_b
	\tau_1 \partial_{\phi}
\ea
where the dynamical phase space variables $\tilde{b}, \tilde{c}, \tilde{p}_b, \tilde{p}_c$
are all only functions of time, and $\tau_i$ are $SU(2)$ generators satisfying
$[\tau_i, \tau_j] = \epsilon_{ij}^k \tau_k$. The relationship
between the triad variables $\tilde{p}_b, \tilde{p}_c$ and
the metric variables $g_{xx}, g_{\Omega \Omega}$ is
given through the general relation
\be
	E^a_i E^b_i =  \det(q) \,q^{ab}
\ee
which explictly implies
\ba
	g_{xx} &=& \frac{\tilde{p}_b^2}{|\tilde{p}_c|} \\
	g_{\Omega \Omega} &=& |\tilde{p}_c|   \,.
\ea
We thus see that the triad component $\tilde{p}_c$ directly
determines the physical radius of the two sphere which
is proportional to $\sqrt{g_{\Omega \Omega}}$. We will
use this fact when interpreting the quantum dynamics.

The classical Hamiltonian consists of a single constraint known as
the Hamiltonian constraint. The explicit form of the Hamiltonian
is given by
\be
	\Ham = \int \frac{-N}{8 \pi G \gamma^2}
		\Big[ 2 \,\tb\, \tc \,\sqrt{\tpc} + (\tb^2 + \gamma^2)
		\frac{\tpb}{\sqrt{\tpc}}\Big]\, dx\, d\Omega \,.
\ee
Performing the spatial integrations we run into a problem
as the integral over $x$ diverges. Following~\cite{Ashtekar:2005qt}, if
we restrict the integration over $x$ to a finite interval $L_0$, the
Hamiltonian becomes
\be\label{Hcl}
	\Ham = \frac{-N}{2 G \gamma^2}
		\Big[ 2 \,b\, c \,\sqrt{p_c} + (b^2 + \gamma^2)
		\frac{p_b}{\sqrt{p_c}}\Big]
\ee
written in terms of rescaled variables
\be \label{untild}
	b = \tb, \qquad c = L_0 \tc, \qquad p_b = L_0 \tpb, \qquad p_c = \tpc
\ee
The Poisson  structure of these phase space variables is given by
the only non-vanishing Poisson brackets
\ba
	\{b, p_b\} &=& G \gamma \\
	\{c, p_c\} &=& 2 G \gamma
\ea
Written in terms of these variables, the metric~\eqref{metric} is given
by
\be\label{triadmetric}
	ds^2 = - N^2 dt^2 + \frac{p_b^2}{|p_c|} \frac{dx^2}{L_0^2} + |p_c| d\Omega^2
\ee

The equations of motion are given as in an ordinary Hamiltonian
dynamical system based on the Hamiltonian~\eqref{Hcl}.
The dynamical behavior of the phase space variables is governed
by Hamilton's equations which give for instance
\be
	\dot{p}_b = \{p_b, \Ham\} = - G \gamma \frac{\partial \Ham}{\partial b}
\ee
and so forth. Additionally, the lapse $N$ appears in the Hamiltonian as
a Lagrange multiplier and this implies that the Hamiltonian must vanish
leading to an additional constraint on the dynamics
\be
	\Ham = 0 \,.
\ee

We will next consider two classes of solutions that will be relevant to the discussion.
The first consists of the usual Schwarzschild interior we are most interested in. The second
corresponds to the Nariai solution which will be of relevance when we discuss the modified quantum
dynamics.

\subsection{Schwarzschild solution}
To arrive at the Schwarzschild interior solution
let us first choose for simplicity the lapse (for $b \neq 0$)
to be given by
\be
	N = \frac{\gamma \sqrt{p_c}}{b}
	\label{eq:lapse1}
\ee
whence the Hamiltonian becomes
\be
	\Ham = \frac{-1}{2 G \gamma}
		\Big[ 2 \, c \,p_c + (b + \gamma^2/b)
		p_b\Big] \,.
\ee
The equations of motion are given by
\ba
	\dot{b} &=& - \frac{1}{2}(b+\gamma^2/b) \\
	\dot{c} &=& -2c\\
	\dot{p}_b &=& \frac{p_b}{2}(1-\gamma^2/b^2) \\
	\dot{p}_c &=& 2 p_c \\
	\Ham &=& 0 \Rightarrow 2c = -\frac{p_b}{p_c}(b + \gamma^2/b) \,.
\ea
The solution to these equations of motion is given by
\ba
	p_b(T) &=&  p_b^{(0)} e^{T}\sqrt{2 m e^{-T}-1} \label{pbcl}\\
	p_c(T) &=& e^{2T} \label{pccl}\\
	b(T) &=& \pm \gamma \sqrt{2m e^{-T}-1} \label{bcl}\\
	c(T) &=& -\gamma m p_b^{(0)} e^{-2T} \label{ccl}\,.
\ea
with $p_b^{(0)}$ a constant of motion.
Transforming to a new time coordinate $T = \log t$, we
have $N dT = \gamma \frac{\sqrt{p_c}}{b} \, \frac{dt}{t} = 1/\sqrt{2m/t-1} \;dt$
and the metric~\eqref{triadmetric} is given by
\be
	ds^2 = - \frac{1}{\frac{2m}{t}-1} dt^2 +
	\bigg(\frac{p_b^{(0)}}{L_0^2}\bigg)^2 \Big(\frac{2m}{t}-1\Big) dx^2 + t^2 d\Omega^2
\ee
If we then identify $p_b^{(0)} = L_0$, we arrive at the Schwarzschild interior
solution~\eqref{Swcl}.

In these  coordinates, the singularity is located at $t = 0$, while
the horizon is at $t = 2m$. We can identify the singularity and horizon
in a coordinate invariant fashion by noting that at the singularity
both triad components $p_c$ and $p_b$ vanish. The horizon then occurs
where $p_b = 0$ and $p_c = 4 m^2$.

Before moving to the second class of solutions let us discuss certain gauge freedoms
of the Schwarzschild solution. Classically there are two gauge freedoms corresponding
to coordinate changes in either $t$ or $x$ variables. A change in the time
coordinate $t$ leads to a change in the lapse leaving all physics unchanged. On the other
hand if we consider a coordinate rescaling of  $x$, $L_0$ is correspondingly rescaled such
that the ratio $dx/L_0$ is left invariant. This
implies that a rescaling of $x$ does not rescale the value of $p_b$. Note that this point
was mistaken in~\cite{Ashtekar:2005qt} where it was mentioned that a rescaling
of $p_b^{(0)}$ can be accommodated by a rescaling in $x$.

In addition to the classical coordinate gauge freedoms, there is an additional
gauge in the loop setup pertaining to the choice in restriction of the spatial
integration leading to the $L_0$ parameter. From the definitions of
$p_b$ and $c$ in~\eqref{untild}, it is easy to see that if we choose a different
size of $L_0$ then this will rescale the value of $p_b$ and $c$ and hence in
terms of the Schwarzschild solution, $p_b^{(0)}$ will rescale accordingly.

The loop quantization is based on the phase-space variables $p_b, p_c, b, c$ all
of which are invariant under a coordinate rescaling of $x$. The quantization
is thus manifestly invariant under the first gauge transformation. However
the quantum effects in principle can depend on the second gauge freedom
which does not appear classically. This will be manifested in the quantum
dynamics by a dependence of the physical results on the parameter $p_b^{(0)}$.
Any dependence of the quantum physics on this gauge freedom implies
the need for additional input into the theory (namely the specification of
the value of $p_b^{(0)}$) which must be done by hand. We will show that
the improved quantization we focus on in section \ref{gooddelta} will be independent of $p_b^{(0)}$ and
hence will not require additional input of its value.

\subsection{Nariai solution}
An additional classical solution that will be of relevance to the discussion
of the effective dynamics is the Nariai solution~\cite{Nariai2,Nariai1}.
It differs from the Schwarzschild solution
in that the two-sphere radius determined from $p_c$ is assumed to be
constant and the model is sourced by a cosmological constant with the possible
inclusion of charge. The assumption that $p_c$ is a constant requires
that $b=0$ and hence the lapse~\eqref{eq:lapse1} used in the Schwarzschild
case becomes degenerate. We must therefore choose another
form of the lapse to generate the Nariai solution.

Choosing the lapse to be equal to one and adding in a cosmological constant $\Lambda$
leads to the Hamiltonian
\begin{align}
      H = \frac{-1}{2 G \gamma^2} \Bigl[
      2 b c \sqrt{p_c} + (b^2 + \gamma^2)\frac{p_b}{\sqrt{p_c}} \Bigr] +
      \frac{\Lambda}{2 G} p_b \sqrt{p_c}.
      \label{lam1}
\end{align}
We are interested in a solution where the radius of the two-sphere is
constant implying $p_c = {\rm const}$, such that $\dot{p}_c =0$. The resulting
equations of motion imply $b=0$. Putting $b=0$ into the equation for
$\dot{b}$ yields the relation between the constant radius of the sphere and
the cosmological constant $p_c = 1/\Lambda$.
The remaining two equations of motion simplify to
\begin{align}
      \dot{c} &= \gamma \Lambda \sqrt{\Lambda} p_b
      \label{lam6}\\
      \dot{p}_b &= \frac{1}{\gamma} \frac{1}{\sqrt{\Lambda}} c
      \label{lam7}
\end{align}
which can be written as a single second order differential equation for
$p_b$  given by
\begin{align}
      \ddot{p}_b = \Lambda p_b.
      \label{lam8}
\end{align}
The general solution is given by
\begin{align}
      p_b = c_1 \cosh(\sqrt{\Lambda}t) + c_2
\sinh(\sqrt{\Lambda}t)
\end{align}
where $c_1$ and $c_2$ are two constants of integration. To obtain the global
cosmological metric of the Nariai Universe we set $c_1  = 1/(\Lambda)$ and
$c_2 = 0$ to get
\begin{align}
      p_b = \frac{1}{\Lambda} \cosh(\sqrt{\Lambda}t).
      \label{lam10}
\end{align}
Hence, we obtained the following solution of the classical field equations
\begin{align}
      ds^2 = -dt^2 + \frac{1}{\Lambda} \cosh^2(\sqrt{\Lambda}t) \frac{dx^2}{L_0^2} +
\frac{1}{\Lambda} d\Omega^2
      \label{lam11}
\end{align}
compare for example~\cite{Bousso:1996pn,Ortaggio:2001af}, which corresponds
to the uncharged Nariai Universe.

Geometrically Nariai type spacetimes can be understood as four dimensional
submanifolds of a flat six dimensional Lorentzian manifold being the product
of two (three dimensional) spaces of constant curvature. If the respective
length scales are denoted by $a$ and $b$ say, the generic Nariai type metric
takes the form
\begin{equation} \label{Nariaimetric}
      ds^2 = a^2 (-d\tau^2 + \cosh^2\tau d\chi^2) + b^2 d\Omega^2.
\end{equation}
For $a^2 = b^2 = 1/\Lambda$, this corresponds to the uncharged Nariai universe,
given by~(\ref{lam11}) upon a rescaling of the time $\tau = t/a$. If $a \neq b$,
the spacetime corresponds to the charged Nariai universe. It should
be noted that the knowledge of the metric alone, does not suffice to distinguish
between the two types of possible charges, electric or magnetic.

\subsection{Radial geodesics}
In order to interpret the quantum dynamics, we will also consider the trajectory
of a radially in-falling test particle.
The geodesics can be derived from the line element
\begin{align}
      ds^2 = -N(t)^2 dt^2 + \frac{p_b^2(t)}{|p_c (t)|} \frac{dx^2}{L_0^2} + |p_c (t)| d\Omega^2.
      \label{r1}
\end{align}
Since we are mainly interested in radial geodesics we henceforth assume
$d\Omega^2 = {\rm constant}$. The geodesic equations are most easily read
off from the Lagrangian
\begin{align}
      -2\mathcal{L} = -N^2 \bigg(\frac{dt}{d\tau}\bigg)^2 +
      \frac{p_b^2}{p_c L_0^2} \bigg(\frac{dx}{d\tau}\bigg)^2
      \label{r2}
\end{align}
where $\mathcal{L}=0$ for a massless particle and $\mathcal{L}=1/2$ for
a massive particle, and $\tau$ refers to the proper time for a massive
particle.  Since the Lagrangian~(\ref{r2}) is independent
of the variable $x$, its conjugate momentum $\pi_x$ is conserved
\begin{align}
      \pi_x = \frac{\partial \mathcal{L}}{\partial \dot{x}} =-
      \frac{dx}{d\tau} \frac{p_b^2}{p_c L_0^2} = {\rm const} =- \mathcal{E}
      \label{r3}
\end{align}
where $\mathcal{E}$
represents the total energy including the gravitational potential energy
for a static observer at infinity.
Since $dx/d\tau=\mathcal{E}p_c L_0^2/p_b^2$ the
Lagrangian can now be written as
\begin{align}
      -2\mathcal{L} =  -N^2 \bigg(\frac{dt}{d\tau}\bigg)^2
      + \frac{p_c L_0^2}{p_b^2} \mathcal{E}^2.
      \label{r4}
\end{align}
Hence, we obtain
\begin{align}
      \bigg(\frac{dt}{d\tau}\bigg)^2 = \Bigl(\frac{p_c L_0^2}{p_b^2} \mathcal{E}^2 +
      2\mathcal{L}\Bigr)\frac{1}{N^2}.
      \label{r5}
\end{align}
In the classical Schwarzschild case, we have $p_c L_0^2 / (N^2 p_b^2)=1$ and
hence the geodesic equation can be understood as a point particle with energy $\mathcal{E}$
moving in an effective potential given by $2 \mathcal{L}/ N^2$. Moreover, the geodesic
equation is simple to interpret because of the simple relation between the coordinate
parameter $t$ and the two-sphere radius determined from $p_c$ since
we have $p_c = t^2$. However, when we come to the quantum dynamics,
the relationship between the coordinate parameter $t$ and $p_c$ will become
complicated and we will no longer be interested in the behavior of the coordinate
parameter $t$ as a function of proper time. We will therefore be most interested
in $p_c(\tau)$ which can be derived from
\be \label{geodesic}
	\frac{d p_c}{d\tau} = \frac{d p_c}{dt} \frac{dt}{d\tau}
\ee
where $d p_c/dt$ is to be determined from the equations of motion
derived from the Hamiltonian, and $dt/d\tau$ is given by~\eqref{r5}.

Classically, for any given energy $\mathcal{E}$, the in-falling particle
reaches the singularity at $p_c=0$ in finite proper time. We will be interested
in whether the quantum dynamics implies that the singularity is never reached
by the in-falling particle. We will see that the loop quantum dynamics imply a
minimum value of the two-sphere radius $p_c$ and hence the solutions
are non-singular with an in-falling particle never reaching the classical singularity.

\section{Effective Loop Quantum Dynamics}

In attempting to ascertain the quantum corrections to the classical behavior
implied by the loop quantization, we will work in an effective semi-classical
description that leads to modifications of the classical Hamiltonian constraint.
The effective description is motivated from the construction of
the quantum Hamiltonian operator, although a more rigorous
understanding of the quantum dynamics would require more sophisticated
machinery. Our aim here is to describe the expected dominant corrections
arising from the quantum theory and as such, there is room for additional
corrections that could modify some of the results presented here.
Evidence that the effective theory we describe provides an accurate
description of the quantum dynamics is given by homogeneous and isotropic
loop quantum cosmology with a massless scalar field with or without a
cosmological constant~\cite{Ashtekar:BBII,Newkplus1,Vandersloot:2006ws}.

The main quantum correction arises from the fact that in the loop quantization
no direct operators exist corresponding to the connection components
$b$ and $c$. Instead, the quantization proceeds  as in loop quantum gravity
by considering holonomies of the connection. The holonomies roughly
consist of exponential terms like $\exp(i b \delta_b)$ with $\delta_b$ corresponding
to the {\em edge length} of the holonomy. The end result in the effective Hamiltonian
is to replace the classical $b$ terms in the Hamiltonian~\eqref{Hcl}, with
$\sin( b \delta_b)/\delta_b$ terms for instance. Thus the $\delta_b, \delta_c$ factors
play a crucial role in determining the effects of the quantum corrections.
The classical limit is recovered by taking the $\delta_b, \delta_c \rightarrow 0$ limits.
The $\delta_b, \delta_c$ parameters, arising from the holonomy edge lengths,
are  a measure of the quantum discreteness arising from the loop
quantization.

In the original loop quantization of the Schwarzschild interior~\cite{Ashtekar:2005qt},
the $\delta_b, \delta_c$ terms were assumed to be a constant value labeled by $\delta$
analogous to the original quantization of loop quantum cosmology where
similar terms appear in the Hamiltonian operator. In~\cite{Ashtekar:BBII}, it
was shown that because of the assumption that $\delta$ is a constant (in loop quantum
cosmology the parameter is labeled $\mu_0$),
loop quantum cosmology does not have the correct semi-classical limit.
The solution to this issue presented in that article is to relax the assumption
that the $\delta$ factors are constant and instead consider them to be  functions of the triad
variables. The precise prescription given leads to the non-constant factor
labeled $\bar{\mu}$ which varies inversely with the scale factor of the universe and
with this behavior the correct semi-classical limit is achieved \cite{Ashtekar:BBII}.

Thus our goal will be to consider the effective dynamics with a quantization scheme
arising from the improved prescription for loop quantum cosmology given
in~\cite{Ashtekar:BBII}. In particular, the $\delta_b, \delta_c$ factors will depend
explicitly on the triad components $p_b$ and $p_c$. We will start however, with
the case of constant $\delta$ since the effective dynamics are exactly solvable
and allows to contrast the modifications from those arising from the new non-constant
$\delta$ quantization. The results for the constant $\delta$ case have been previously
derived in~\cite{Modesto:2006mx}. We merely repeat the results that
are relevant to our discussion.

\subsection{Constant \boldmath{$\delta$} Quantum Hamiltonian}
\label{constantd}
With the considerations of the previous discussion
the quantum modifications can be understood as leading
to an effective Hamiltonian with the holonomy parameter
$\delta$ playing a crucial role in determining the magnitude of
the corrections.
The effective Hamiltonian constraint reads~\cite{Modesto:2006mx}
\begin{align}
      \Ham_{\delta} = -\frac{N}{2 G \gamma^2} \Bigg[
      2 \frac{\sin  b \delta}{\delta} \frac{\sin  c \delta}{\delta} \sqrt{p_c} +
      \Big(\frac{\sin^2\negmedspace b \delta}{\delta^2} + \gamma^2\Big)
      \frac{p_b}{\sqrt{p_c}} \Bigg].
      \label{a9}
\end{align}
It is easy to see that in the limit $\delta \rightarrow 0$, the classical
Hamiltonian~\eqref{Hcl} is recovered. In analogy to the classical case,
we choose $N=\gamma \sqrt{p_c} \delta/(\sin  b \delta)$ and the equations
for the pairs $(b,p_b)$ and $(c,p_c)$ decouple to two independent pairs of
differential equations
\begin{align}
      \dot{c}  &= 2\gamma G
      \frac{\partial\Ham_\delta}{\partial p_c} = -2 \frac{\sin  c \delta}{\delta}
      \label{a10}\\
      \dot{p_c}  &= -2\gamma G
      \frac{\partial\Ham_\delta}{\partial c} = 2 p_c \cos  c \delta
      \label{a11}\\
      \dot{b}  &= \gamma
      \frac{\partial\Ham_\delta}{\partial p_b}
      = -\frac{1}{2}\Bigl(\frac{\sin  b \delta}{\delta} +
      \frac{\gamma^2 \delta}{\sin b \delta}\Bigr)
      \label{a12}\\
      \dot{p_b}  &= -\gamma
      \frac{\partial\Ham_\delta}{\partial b}
      = \frac{1}{2}\cos  b \delta \Bigl(1-\frac{\gamma^2 \delta^2}
      {\sin^2\negmedspace b \delta}\Bigr) p_b.
      \label{a13}
\end{align}
The equation of motion for $c$ can be integrated to give
\begin{align}
      c(T) = \frac{1}{\delta} \arctan\Bigl(\mp \gamma m p_b^{(0)}
      \frac{\delta}{2} e^{-2T} \Bigr)
      \label{a14}
\end{align}
where the constant of integration has been fixed so that the limit
of the effective $c(T)$ as $\delta \rightarrow 0$ coincides with
the classical expression (\ref{ccl}). Next, the equations for $p_c$
can be solved
\begin{align}
      p_c(T) = e^{2T} + \gamma^2 m^2 (p_b^{(0)})^2 \frac{\delta^2}{4} e^{-2T}.
      \label{a15}
\end{align}
Also the equation for $b$ can be integrated and yields
\begin{align}
      b(T) = \frac{1}{\delta}\arccos\Biggl[ \sqrt{1+\gamma^2\delta^2}
      \tanh\biggl\{\sqrt{1+\gamma^2\delta^2}\Bigl(\frac{T-\log 2m}{2}-
      \log\frac{\gamma\delta}{2}\Bigr)\biggr\}\Biggr].
      \label{a16}
\end{align}
Finally, $p_b(T)$ can be obtained from the vanishing of the
Hamiltonian~\eqref{a9} which gives
\be
	p_b =  -\frac{\sin  b \delta}{\delta} \frac{\sin  c \delta}{\delta}
	\frac{2 p_c}{\sin^2\negmedspace b \delta / \delta^2 + \gamma^2}.
\ee

\begin{figure}[!ht]
\noindent
\begin{minipage}[h]{.48\linewidth}
\centering\epsfig{figure=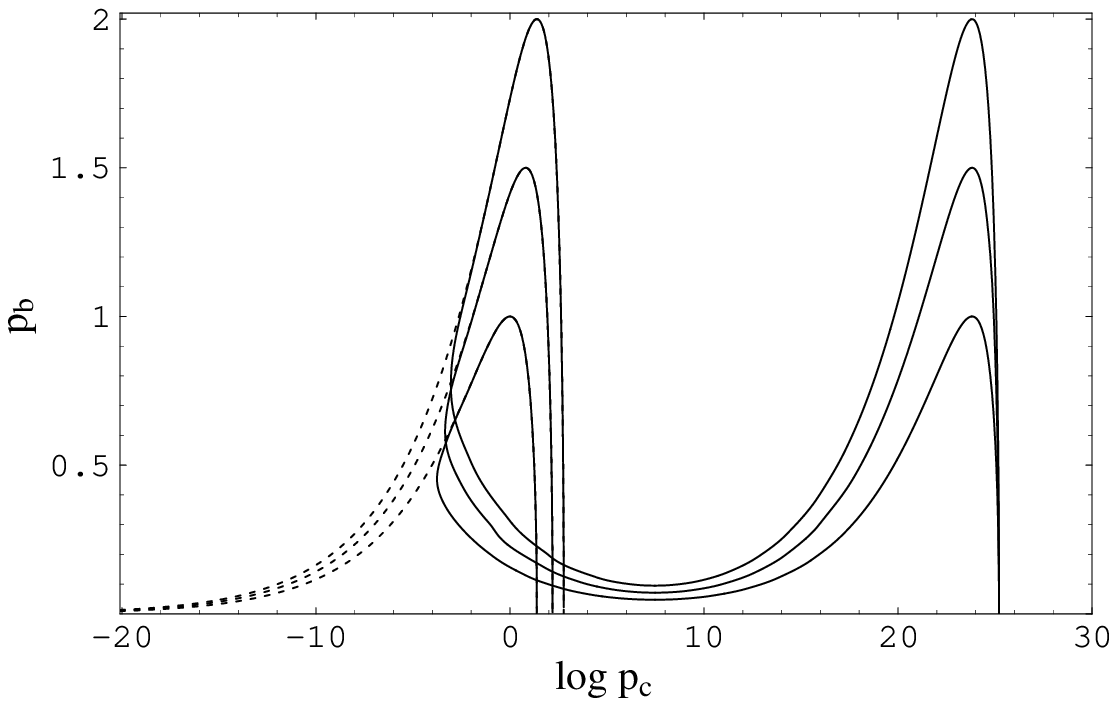,width=\linewidth}
\end{minipage}\hfill
\begin{minipage}[h]{.48\linewidth}
\centering\epsfig{figure=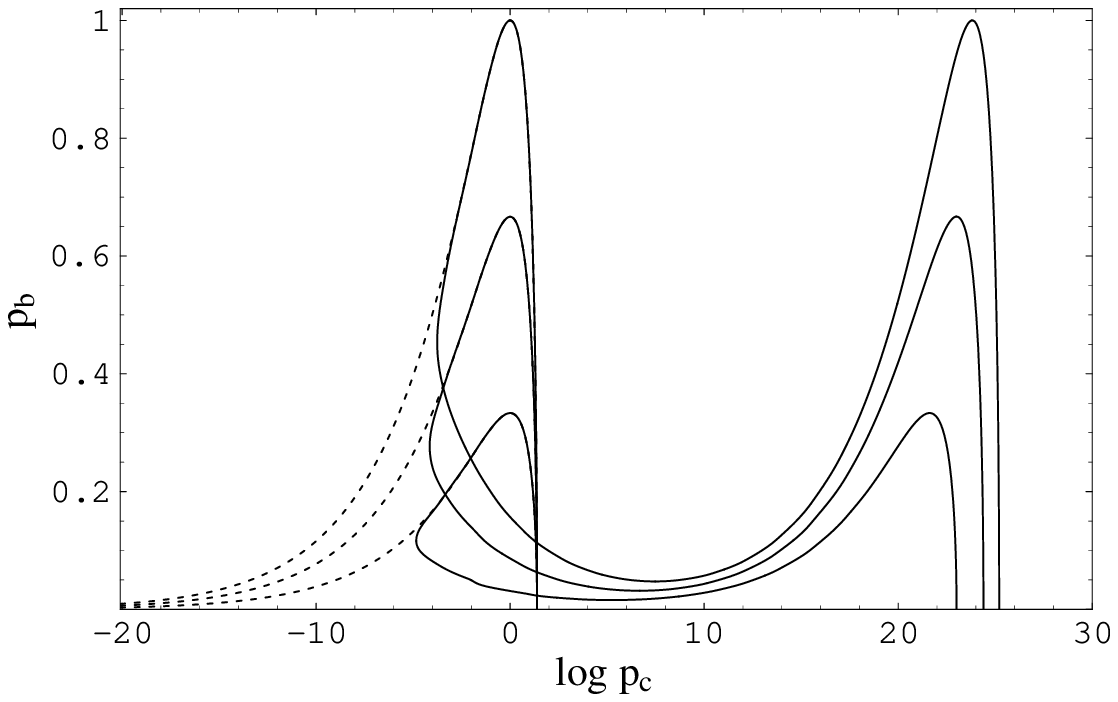,width=\linewidth}
\end{minipage}
\caption{These figures show the effective plots of $p_b = p_b (p_c)$ for the case
with $\delta = const$. The dashed lines
represent the classical solution while the solid lines correspond to the solution of
the effective quantum theory with $\delta=0.1$. Clearly, the initial singularity is
removed and the spacetime contains a second black hole.
The three plots on the left show the solution for varying
mass parameter, $m = 1; 1.5; 2$ respectively. On the right $p_b^{(0)}$ is varied in the
three plots, $p_b^{(0)} = 1; 2/3; 1/3$ respectively. One clearly notes (left) that the
mass parameter $m$ controls the mass of the first (classical) black only, while (right)
the parameter $p_b^{(0)}$ controls the mass of the second black hole.}
\label{fig:const1}
\end{figure}

>From the form of $b(T)$, it is evident that the evolution ends
at a minimum value  $T_{\rm min}$ where the
absolute value of the argument of the $\arccos$ function equals one.
Moreover, from the form of $p_c(T)$ it is evident that the group orbits
also have a minimal value at $T=T_{\rm min}=\log(\gamma m p_b^{(0)} \delta/2)/2$,
with $p_c$ reaching its maximum value at the classical horizon and at
$T_{\rm min}$.
This minimal two-sphere radius is given from $p_c(T_{\rm min})=\gamma m p_b^{(0)} \delta$,
which depends on the mass of the black hole and also on $p_b^{(0)}$.
Figure~\ref{fig:const1} shows
the plots of $p_b = p_b (p_c)$ for various values of $m$ and $p_b^{(0)}$.
The singularity is thus avoided through a bounce in the two-sphere radius and
the quantum dynamics matches two black holes together though in a generically
non-symmetric value. A symmetric solution can be constructed by fine tuning
the value of $p_b^{(0)}$.

As we have seen in the figures, the mass parameter $m$ controls the mass of the first black
hole whereas $p_b^{(0)}$ controls the mass of the second black hole. This suggests that
in the effective quantum theory, $p_b^{(0)}$ would not be regarded as a meaningless
constant of integration or a certain choice of gauge. For the constant $\delta$ quantization
of the Schwarzschild interior spacetime this parameter acquires an important physical
meaning, and two different choices correspond to different effective quantum theories.
Any physical results derived from the constant $\delta$ quantization
must include a prescription for specifying the value of $p_b^{(0)}$ which
has to be input by hand.

\subsection{Improved Quantum Hamiltonian} \label{gooddelta}
We now wish to relax the assumption that the $\delta$ parameters are constant.
In accordance with the results from loop quantum cosmology~\cite{Ashtekar:BBII},
we will assume that $\delta_b, \delta_c$ depend on the triad variables
$p_c$ and $p_b$. The method to constrain the exact dependence
relies on the fact that the holonomies used to construct the quantum
Hamiltonian operator form closed loops. The behavior of $\delta_b$ and
$\delta_c$ can then be constrained if we demand that the physical area
of the closed loop be equal to the minimum area gap predicted by loop
quantum gravity
\be
	A_{\rm min} = \Delta = 2 \sqrt{3} \pi \gamma \,\lp^2.
\ee

This procedure can be executed in different possible ways in the Kantowski-Sachs
model considered here, whereas homogeneous isotropic models lead to a unique
choice. We discuss in this section what we feel to be the most geometrically
natural scheme. We will show in the section, that this scheme has the
advantage that the resulting dynamics is independent of the $p_b^{(0)}$
parameter. We will discuss an alternative
scheme in appendix \ref{Alt}.

The procedure considered in this section amounts to evaluating the classical area
of the holonomy loops and constraining the area based on this. For instance
if we consider the holonomy loop in the $x, \theta$ plane, the classical physical
area is given by
\be
	A_{x \theta} = \delta_b \delta_c  \,p_b
\ee
which is evident from the form of the metric~(\ref{triadmetric}).
For the holonomies on the two-sphere, the loop does not close
as discussed in~\cite{Ashtekar:2005qt}, but we can assign
an effective area given by
\be
	A_{\theta \phi} = \delta_b^2 p_c \,.
\ee
If we  constrain these areas to be equal to $A_{\rm min}$ we have the following
behavior
\ba \label{deltagood}
	\delta_b &=& \frac{\sqrt{\Delta}}{\sqrt{p_c}}\\
	\delta_c &=& \sqrt{\Delta} \frac{\sqrt{p_c}}{p_b} \,.
\ea
With this, the effective Hamiltonian constraint reads
\begin{align}
      \Ham_{\delta} = -\frac{N}{2 G \gamma^2} \biggl[
      2 \frac{\sin b \delta_b}{\delta_b} \frac{\sin \delta_c c}{\delta_c} \sqrt{p_c} +
      \Bigl(\frac{\sin^2\negmedspace b \delta_b}{\delta_b^2} + \gamma^2\Bigr)
      \frac{p_b}{\sqrt{p_c}} \biggr].
      \label{d1}
\end{align}
Once again, in the limit $\Delta\rightarrow 0$ the effective Hamiltonian~(\ref{d1}) reduces
to the classical Hamiltonian~(\ref{Hcl}).

We can now derive the equations of motion as in the previous case.
Choosing the lapse function to be
$N=\gamma \sqrt{p_c} \delta_b/(\sin b \delta_b)$, the
resulting equations of motion are given by
\ba
      \dot{c} &=&
      -\frac{\sin c \delta_c}{\delta_c} -\frac{1}{2}\frac{\sin\delta_b b}{\delta_c}
      -c \cos c \delta_c \nonumber \\
      &&+ \frac{1}{2}\frac{\delta_b}{\delta_c} b
      \cos b \delta_b
      \Bigl(1-\frac{\gamma^2 \delta_b^2}{\sin^2\negmedspace b \delta_b}\Bigr)
      +\frac{\gamma^2}{2}\frac{\delta_b^2}{\delta_c}\frac{1}{\sin b \delta_b}
      \label{d3}\\
      \dot{p_c}  &=& 2 p_c \cos c \delta_c
      \label{d4}\\
      \dot{b} & =&
      -\frac{1}{2}\Bigl(\frac{\sin  b \delta_b}{\delta_b} +
      \frac{\gamma^2 \delta_b}{\sin b \delta_b}\Bigr) -
\frac{\sin c \delta_c}{\delta_b}
      + \frac{\delta_c}{\delta_b} c \cos c \delta_c
      \label{d5}\\
      \dot{p_b} & =&
      \frac{1}{2}\cos b \delta_b \Bigl( 1 - \frac{\gamma^2 \delta}
      {\sin^2\negmedspace b \delta_b}\Bigr) p_b.
      \label{d6}
\ea

Before considering numerical solutions to these equations, we can
get an idea on where the quantum modifications should be appreciable.
It is evident from the effective Hamiltonian, the the modifications arise
when either $ c \delta_c$ or $ b \delta_b$ are of the order of one.
Near the classical singularity we have $p_b \rightarrow 0, p_c \rightarrow 0,
b \rightarrow \pm \infty$ and $c \rightarrow -\infty$ in such
a way that $ b \delta_b$ and $ c \delta_c$ blow up, hence we expect
the quantum corrections to become important near the singularity.
If we also consider the horizon, we have $p_b \rightarrow 0, p_c \rightarrow (2m)^2,
b \rightarrow \pm 0$ and $c \rightarrow -\gamma m p_b^{(0)} / (2m)^2$
such that $ b \delta_b$ is small, but $ c \delta_c$ diverges. Thus we can
expect quantum corrections also as the classical horizon is approached.
Therefore, the picture that arises is a region in the interior where
classical behavior is recovered with quantum effects near the classical singularity and also
near the classical horizon. This behavior is borne out in the numerical simulations
which we now discuss. We focus in this article on the effects near the singularity and will
comment later on the horizon effects. Accordingly, in the numerical simulations
we will specify initial conditions starting in the interior region where the quantum effects
are negligible and evolve towards the singularity.

\begin{figure}[!ht]
\noindent
\begin{minipage}[h]{.48\linewidth}
\centering\epsfig{figure=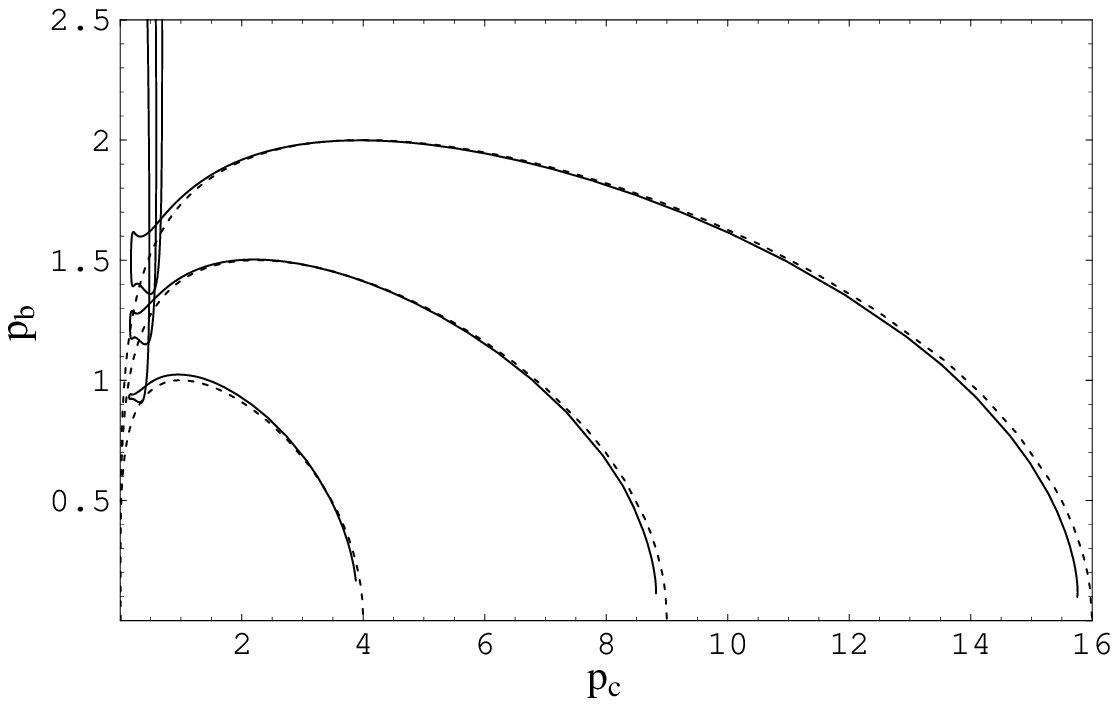,width=\linewidth}
\end{minipage}\hfill
\begin{minipage}[h]{.48\linewidth}
\centering\epsfig{figure=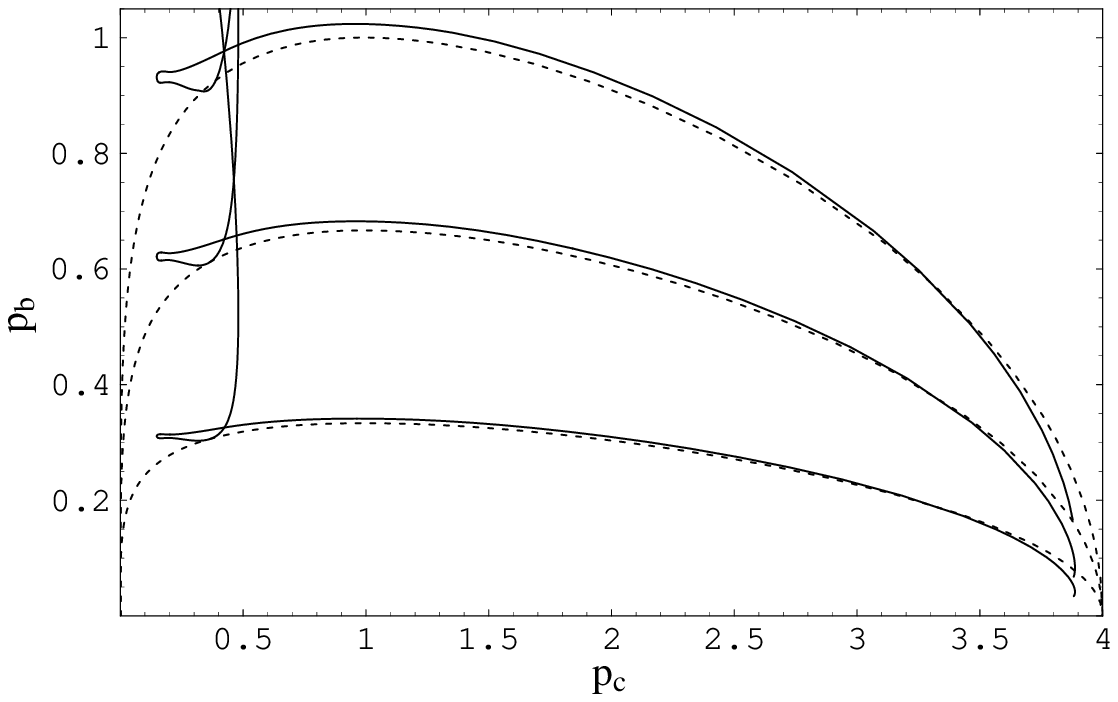,width=\linewidth}
\end{minipage}
\caption{These figures show the effective plots of $p_b = p_b (p_c)$ for the case where
$\delta_b, \delta_c$ are not constants. The dashed lines
represent the classical solution while the solid lines correspond to the solution of
the effective quantum theory. The initial singularity is removed, however, not with a
second black hole. The three plots on the left show the solution for varying
mass parameter, $m = 1; 1.5; 2$ respectively. On the right $p_b^{(0)}$ is varied in the
three plots, $p_b^{(0)} = 1; 2/3; 1/3$ respectively. One verifies that this quantization
scheme does not depend on the choice of $p_b^{(0)}$ and thus the effective theory
is independent of this gauge freedom.}
\label{fig:deltawell1}
\end{figure}

To solve the equations of motion numerically, we start at some initial
point near the horizon with initial conditions $p_c(T_0), p_b(T_0), b(T_0)$ that
match the classical solutions (\ref{pccl},\ref{pbcl},\ref{bcl}) for some initial $T_0$
and the constraint $\Ham_{\delta} = 0$ is then solved to get $c(T_0)$. The coupled
differential equations of motion are then solved for the approach towards
the classical singularity ($T = - \infty$). The results of the simulations
for $p_b$ as a function of $p_c$ appear in figure \ref{fig:deltawell1}
for various initial conditions corresponding to different values of the classical
black hole mass $m$ as well as different values of $p_b^{(0)}$.
The results all indicate that the classical singularity is resolved
with the two-sphere radius $p_c$ being bounded from below.
The second figure of \ref{fig:deltawell1} indicates
that choosing different values of $p_b^{(0)}$ only rescale
$p_b(T)$ as in the classical case, and thus the effective dynamics predicted
is insensitive to this classical gauge choice as opposed to the constant
$\delta$ case.

\begin{figure}[!ht]
\noindent
\begin{minipage}[h]{.48\linewidth}
\centering\epsfig{figure=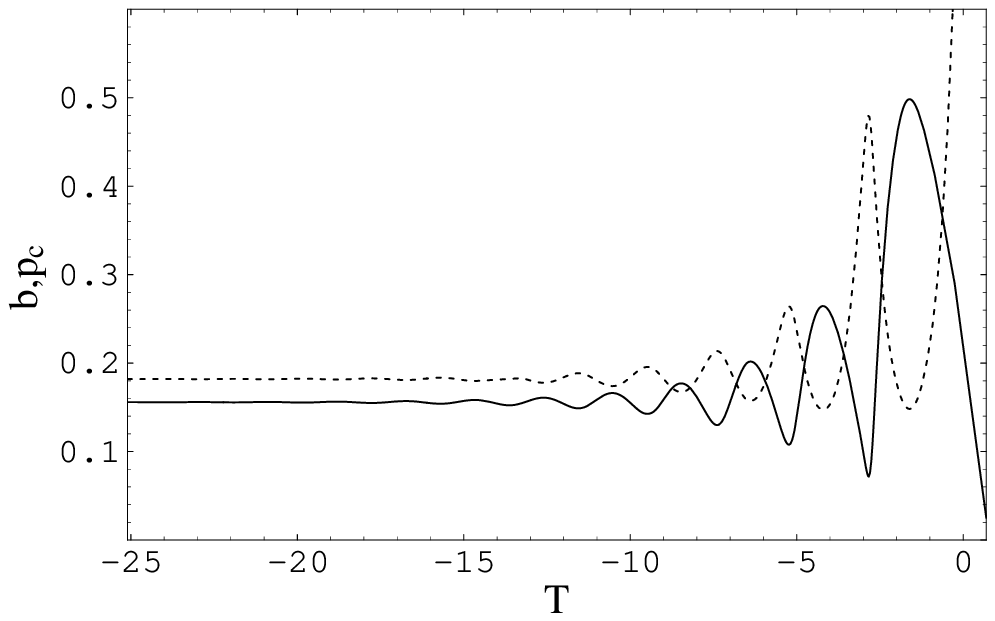,width=\linewidth}
\end{minipage}\hfill
\begin{minipage}[h]{.48\linewidth}
\centering\epsfig{figure=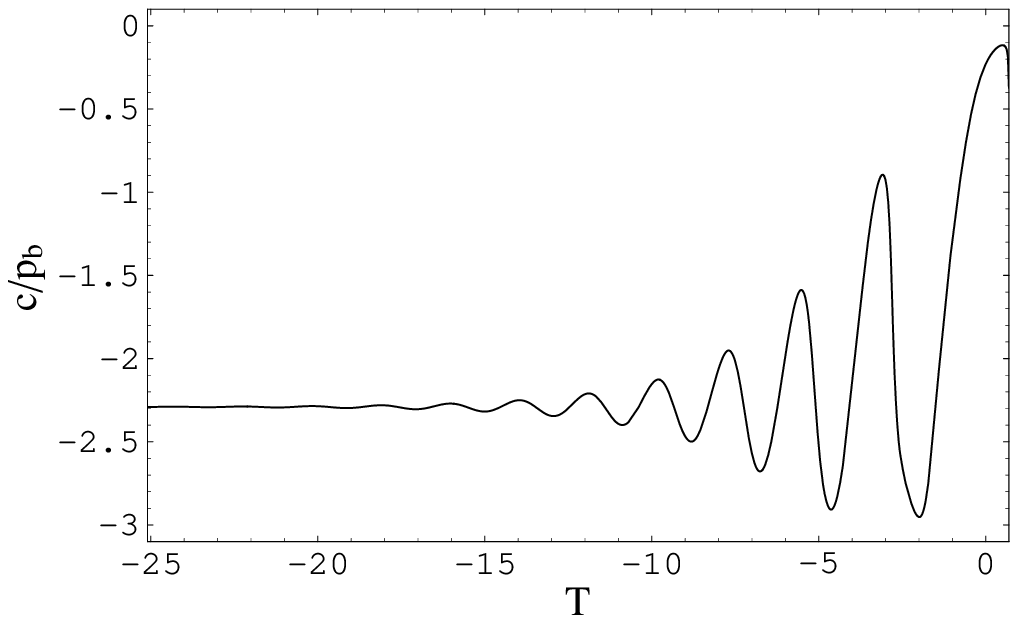,width=\linewidth}
\end{minipage}
\caption{In the left figure $b(T)$ (solid) and $p_c(T)$ (dashed) are plotted.
Both tends towards constants as $T\rightarrow -\infty$. The right figure show
the ratio $c(T)/p_b(T)$ also tends to a constant asymptotically. This asymptotic behavior
is indicative of a Nariai type metric of the form given in equation  \eqref{Nariaimetric}.}
\label{fig:deltawell2}
\end{figure}

Unlike the constant $\delta$ case of section \ref{constantd},
the solutions do not match up to a separate black hole solution.
Instead, the solutions asymptote to a Nariai type solution for
$T \rightarrow - \infty$.
The solution is characterized by
\begin{alignat}{2}
      b &= \bar{b}, & \qquad p_b &= \bar{p}_b\, e^{-\alpha T},
      \nonumber \\
      c &= \bar{c}\, e^{-\alpha T}, & \qquad p_c &= \bar{p}_c,
      \label{d7}
\end{alignat}
where the barred quantities and $\alpha$ are constants.
Figure \ref{fig:deltawell2} shows this behavior
with plots of $p_c, b$ and the ratio of $c/p_b$ as functions of time.
It is clear that $p_c$ undergoes damped oscillations settling in
to a fixed finite value as $T$ goes to minus infinity. Similarly the ratio
of $c/p_b$ settles into a finite value given by $\bar{c}/\bar{p}_b$
as expected from a Nariai type solution.

>From the equations of motion we can determine the asymptotic values
of the constants $\bar{p}_c, \bar{b}, \alpha,$ and $\bar{c}/\bar{p}_b$.
Assuming $p_c=\bar{p}_c$ is a constant, Eq.~(\ref{d4}) implies
\begin{align}
      \cos c \delta_c = \cos\Bigl(\sqrt{\Delta}\frac{\sqrt{\bar{p}_c}}{\bar{p}_b}\bar{c}
      \Bigr) = 0.
      \label{d8}
\end{align}
\begin{align}
      \cos\Bigl(\sqrt{\Delta}\frac{\sqrt{\bar{p}_c}}{\bar{p}_b} \bar{c} \Bigr) = 0,
      \qquad
      \frac{\bar{c}}{\bar{p}_b} = -\frac{\pi}{2 \sqrt{\Delta \bar{p}_c}},
      \label{d9}
\end{align}
which also yield $\sin \bar{\delta}_c \bar{c} = -1$.
Assuming furthermore that $b = \bar{b}$ is constant,
the resulting equation of motion for $\dot{b}$, namely Eq.~(\ref{d5}) yields
\begin{align}
      2\sin\Bigl(\sqrt{\Delta}\frac{\bar{b}}{\sqrt{\bar{p}_c}}\Bigr)-
      \sin\Bigl(\sqrt{\Delta}\frac{\bar{b}}{\sqrt{\bar{p}_c}}\Bigr)^2 =
      \frac{\Delta \gamma^2}{\bar{p}_c}.
      \label{d10}
\end{align}
The same equation is obtained by evaluating the vanishing of the Hamiltonian
constraint~(\ref{d1}).
Next,  Eq.~(\ref{d6}) implies that $p_b(T) = \bar{p}_b e^{-\alpha T}$  and
fixes the constant $\alpha$ to be
\begin{align}
      \alpha =
      -\cos\Bigl(\sqrt{\Delta}\frac{\bar{b}}{\sqrt{\bar{p}_c}}\Bigr)
      + \cot\Bigl(\sqrt{\Delta}\frac{\bar{b}}{\sqrt{\bar{p}_c}}\Bigr).
      \label{d11}
\end{align}
Finally, the above equations Eqs.~(\ref{d9})--(\ref{d11}) can be inserted
in the equation of motion for $c$ leading to
\begin{align}
      -2 + \sin\Bigl(\sqrt{\Delta} \frac{\bar{b}}{\sqrt{\bar{p}_c}} \Bigr)
      -\Bigl(\frac{\pi}{2}+\sqrt{\Delta} \frac{\bar{b}}{\sqrt{\bar{p}_c}} \Bigr)
      \biggl( \cos\Bigl(\sqrt{\Delta}\frac{\bar{b}}{\sqrt{\bar{p}_c}}\Bigr)
      - \cot\Bigl(\sqrt{\Delta}\frac{\bar{b}}{\sqrt{\bar{p}_c}}\Bigr) \biggr) = 0.
      \label{d12}
\end{align}
For given value of $\Delta$ this equation can be solved for  the value
of the ratio $\bar{b}/\sqrt{\bar{p}_c}$. A find root algorithm is necessary since an
analytical solution cannot be obtained. The ratio $\bar{b}/\sqrt{\bar{p}_c}$
then specifies the constant in the exponent $\alpha$ by Eq.~(\ref{d11}).
Moreover, with $\gamma$, $\Delta$ and $\bar{b}/\sqrt{\bar{p}_c}$ given,
the relation~(\ref{d10}) can be used to extract the value of $\bar{p}_c$
which in turn yields the value of $\bar{b}$. Finally, Eq.~(\ref{d9}) fixes
the ratio $\bar{c}/\bar{p}_b$. However, there is a remaining scaling freedom
in $\bar{c}$ and $\bar{p}_b$ since only their ratio is specified. Note that
this implies that rescalings in $\bar{p}_b$ have to be compensated by
simultaneous rescalings in $\bar{c}$.

>From the previous equations it is clear that the asymptotic value of
$\bar{p}_c$ depends only on the constants $\Delta$ and $\gamma$
as
\be
	\bar{p}_c = \gamma^2 g(\Delta)
\ee
where $g(\Delta)$ is some function determined from the solutions
of equations~\eqref{d12} and~\eqref{d10}. For the natural choice
$\Delta=2\sqrt{3}\pi\gamma \lp^2$ the following
asymptotic values are obtained
\begin{alignat}{3}
      \bar{b} &\approx 0.156, & \qquad \bar{p}_c &\approx 0.182\, \lp^2, \qquad \alpha &\approx 0.670
      \nonumber \\
      \bar{c}/\bar{p}_b &\approx -2.290 \, \mplanck^2, &\qquad \bar{N} &\approx 0.689, &
      \label{d13}
\end{alignat}
where $\bar{N}$ is the asymptotic value of the lapse which also behaves as a constant.
These values agree with the asymptotic region of the numerical solution
that we studied, see in particular figure~\ref{fig:deltawell2}.

\begin{figure}[!ht]
\noindent
\begin{minipage}[h]{.48\linewidth}
\centering\epsfig{figure=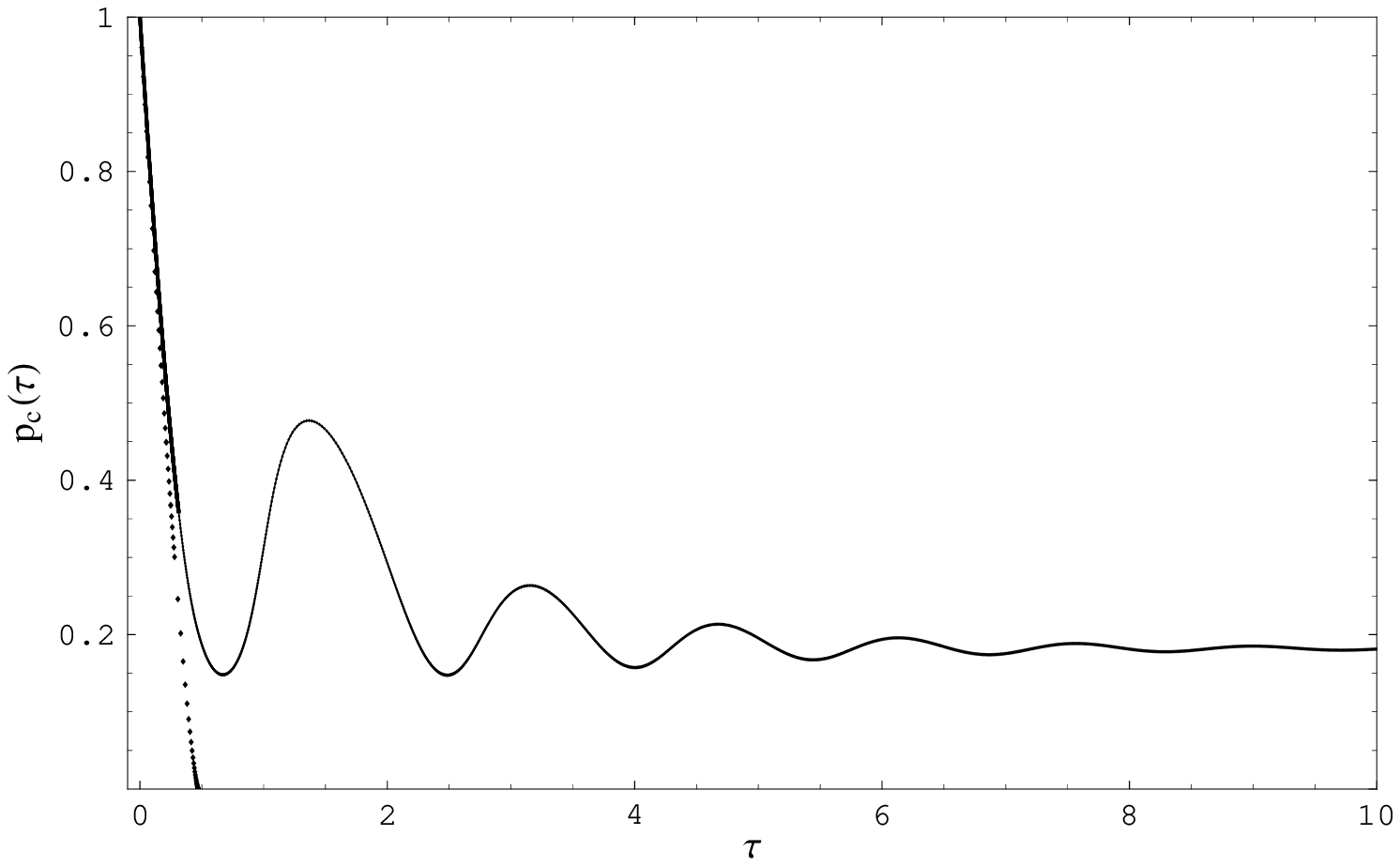,width=\linewidth}
\end{minipage}\hfill
\begin{minipage}[h]{.48\linewidth}
\centering\epsfig{figure=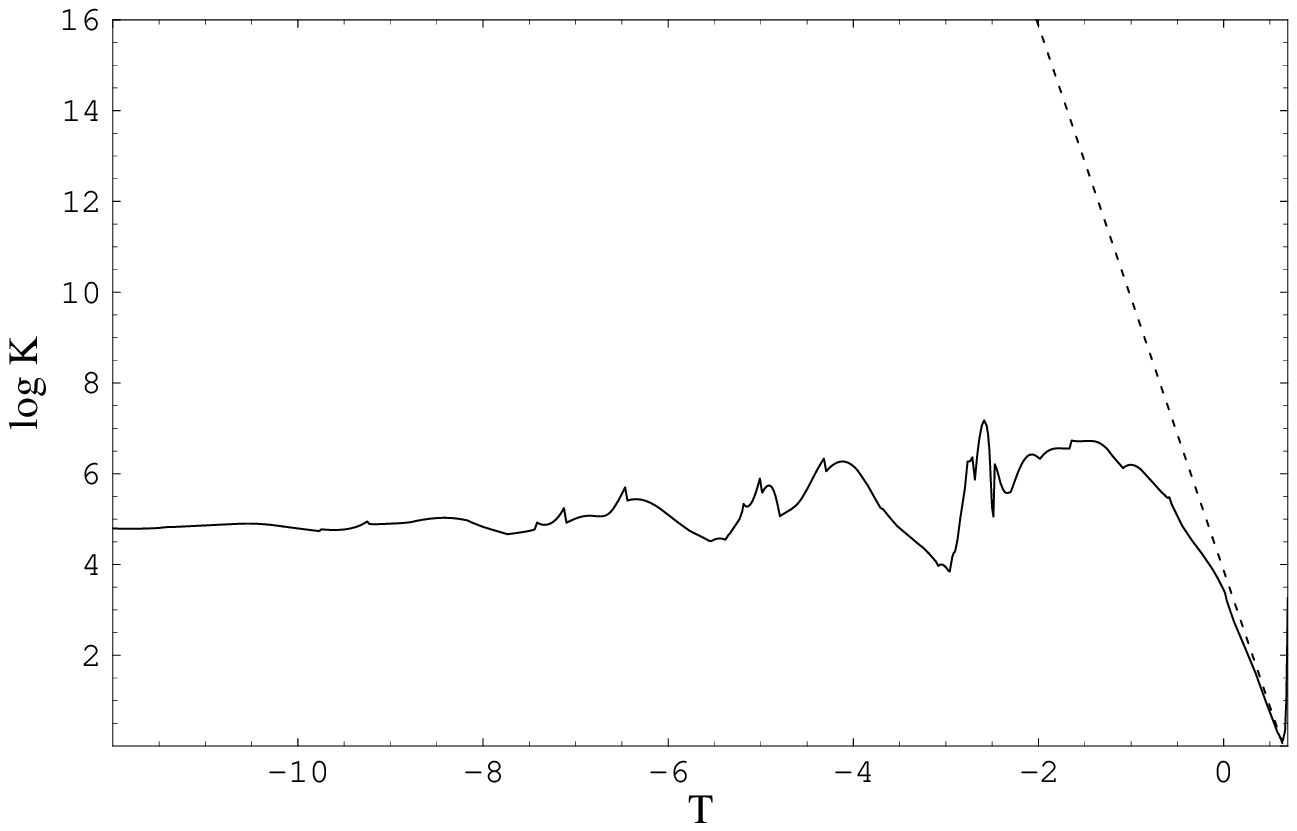,width=\linewidth}
\end{minipage}
\caption{Left: $p_c$ as a function of proper time of radially in-falling geodesic.
The black dashed line represents the classical behavior indicating
that the particle reaches the singularity in finite proper time. The
solid line is the geodesic inside the quantum black hole. The particle
approaches the singularity but undergoes damped oscillations at a fixed two-sphere
radius eventually settling in at a fixed value $\bar{p}_c$. For the natural values of the parameters
$\gamma, \Delta$, the two-sphere radius is on the order of the Planck length.
Right: The logarithm of the Kretschmann invariant as a function of the
time coordinate $T$ is plotted. The dashed line is the classical value which
behaves as $\exp(-6T)$ and diverges at the singularity $T=-\infty$. The solid
line represents the quantum value which asymptotes to a fixed value which
again indicates a resolution of the singularity. The numerical asymptotic values obtained
for $p_c$ and the Kretschmann invariant $K$ are $p_c \approx 0.182\, \lp^2$ and
$K \approx 124.35\, \mplanck^4$ which are both independent of the black hole mass.}
\label{fig:geodesicI}
\end{figure}

The fact that $p_c$ asymptotes to a fixed value on the order of the Planck
length squared implies that the radius of the two-sphere asymptotes to
a value on the order of the Planck length. Furthermore this value is independent
of the mass of the black hole. The interpretation
of the effective dynamics becomes clear if we consider the radially
in-falling test particle using equation~\eqref{geodesic}. A plot
of the numerical integration of $p_c(\tau)$ is shown in figure~\ref{fig:geodesicI}.
An in-falling particle classically would reach the singularity at $p_c = 0$ in
finite proper time as evidenced by the solid black line, however
as a result of the quantum effects, the particle oscillates about the fixed
value of $\bar{p}_c$. The particle thus settles in at a fixed Planckian radius.
The geodesic reaches this asymptotic value in infinite proper time and hence
we can say that the singularity is truly absent in the quantum modified spacetime.
The fate of an in-falling test particle is also of interest in the context of the information loss paradox. The
 results are suggestive that the black hole can store information infinitely long, and strictly speaking no information can ever be lost.
However, once the thermal black hole radiation is taken into account, things are still more involved. In this respect, however, we would like to point out that the improved quantization scheme also leads to quantum modification of the horizon.
In the context
of Euclidean quantum gravity, it has been argued in~\cite{Hawking:2005kf} that indeed horizon modification might resolve the paradox.

We have also considered the behavior of the Kretschmann invariant, the square of the
Riemann tensor $K = R_{abcd} R^{abcd}$.  For the metric~(\ref{triadmetric}), written
out explicitly, this yield a rather complicated expression. However, plotting this
quantity for our numerical solution also shows that the geometry of the spacetime
changes significantly, from the classical Schwarzschild like behavior where
$K = 48m^2 \exp(-6T)$ (which diverges at $T=-\infty$) towards a space of constant curvature,
as expected for the Nariai type metrics whose form is given by
\be
	K = 4 \bigg( \frac{\alpha^4}{\bar{N}^2} + \frac{1}{\bar{p}_c^2} \bigg)\,.
\ee
Plugging in the theoretical values in~\eqref{d13} gives an expected
value
\be
	K \approx 124.36 \,\mplanck^4
\ee
which is in good agreement with the numerical results.
Furthermore this asymptotic value  is independent of the black hole mass.
We find that the quantum effects become appreciable when the Kretschmann curvature
invariant approaches the Planck scale.
Figure~\ref{fig:geodesicI} shows a plot indicating this behavior.

\begin{figure}[!ht]
\noindent
\begin{minipage}[h]{.48\linewidth}
\centering\epsfig{figure=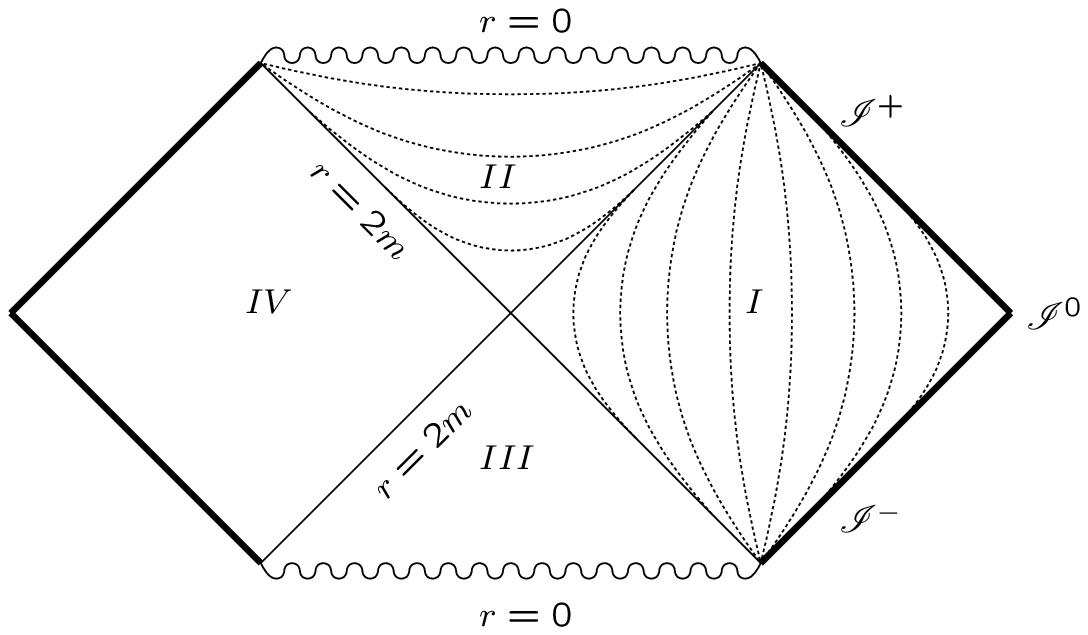,width=\linewidth}
\end{minipage}\hfill
\begin{minipage}[h]{.48\linewidth}
\centering\epsfig{figure=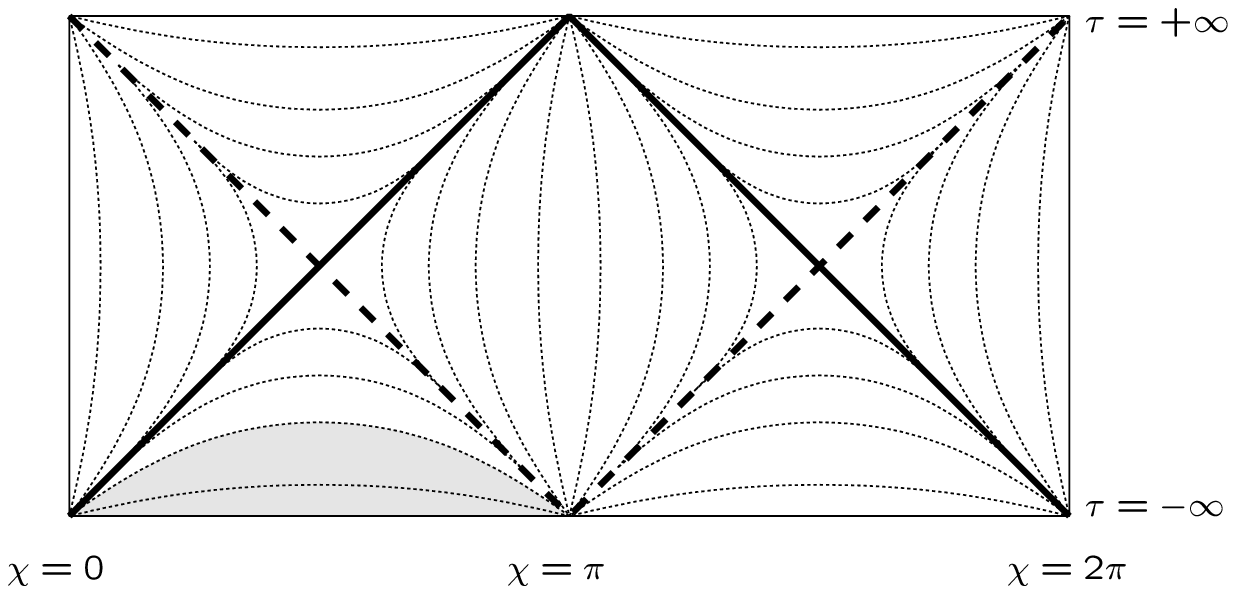,width=\linewidth}
\end{minipage}
\caption{The left figure shows the Carter-Penrose diagram for the Schwarzschild
         spacetime. $\scri^{+}$ and $\scri^{-}$ denote future and past null
         infinity, respectively. The usual Schwarzschild metric in curvature
         coordinates only covers region $I$. The Kantowski-Sachs form of the
         line element covers region $II$. The dotted lines represent the
         $r={\rm constant}$ lines in the respective regions.
	 The right figure shows the Carter-Penrose diagram of the non-singular
         charged Nariai universe. The $\chi=0$ and $\chi=2\pi$ lines are
         identified and the dotted lines stand for the constant two-sphere radii. The solid and dashed diagonal lines represent the disconnected
         future and past event horizon, respectively. Every point of this
         diagram represents a sphere of the same area. For later purpose
         we have indicated a shaded region.}
\label{figa}
\end{figure}

\begin{figure}[!ht]
\noindent
\begin{minipage}[h]{.30\linewidth}
\centering\epsfig{figure=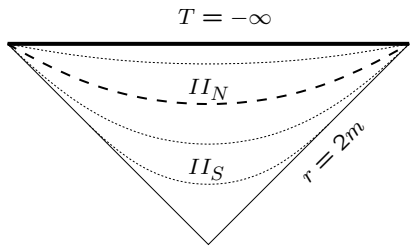,width=\linewidth}
\end{minipage}
\caption{In region $II$ we have indicated a specific $r={\rm constant}$ line. This
         lines separates the region with decreasing group orbits ($II_S$) from the
         region where the 2-spheres have (effectively) constant volume ($II_N$), which defines the
         Nariai spacetime.}
\label{figc}
\end{figure}

The quantum effects removing the singularity can best be visualized by Carter-Penrose
diagrams, see e.g.~\cite{HawkingEllis}. The classical diagrams for
both the Schwarzschild and Nariai spacetimes are given in figure~\ref{figa}.
The quantum effects remove the classical Schwarzschild singularity
shown as the wavy line at the top of region II in the Carter-Penrose diagram and glue
the diagram to the shaded region of the Nariai diagram. The resulting diagram
of the interior is shown in figure~\ref{figc}.

\section{Discussion}

At the level of the effective description we have presented, we have been
able to answer the questions posed at the beginning, namely what replaces
the singularity in the Schwarzschild interior. In all of the quantization schemes
we have discussed, singularity avoidance arises naturally thus providing
further evidence that loop quantum gravity successfully resolves the classical
black hole singularity. We have seen two possible replacements for
the singularity, one in which quantum effects connect two separate (non-identical)
black holes resembling a wormhole-like solution. The other
solution predicting repulsive gravitational effects leading to an equilibrium
two-sphere radius to which an in-falling particle would asymptote.

The results are suggestive but can not be considered as a proof that the LQG
resolves the black hole singularity. The effective equations considered here are not
rigorously derived from the quantum theory and thus can only be considered
approximations of the quantum dynamics. Since understanding
the quantum dynamics is difficult, we do not have a complete handle on when
the approximation might break down. As was mentioned however, the effective
equations have approximated the quantum dynamics in the models where both
can be developed completely, namely the homogeneous and isotropic models of LQC sourced
with a massless scalar field.

Some initial attempts at constructing semi-classical states for the constant $\delta$ quantization
have be made in \cite{Cartin:2006yv} and \cite{Rosen:2006bga} and it is worthwhile to compare
our results with the previous work. In the works cited, numerical instabilities of the difference
equation are found which bring into question whether any effective picture is valid near the singularity.
In \cite{Cartin:2006yv} it is argued that no proper semi-classical states exist unless the Immirzi
parameter $\gamma$ is less the the accepted value in the LQG literature raising a further
question. In this article, our aim has been to probe the improved, non-constant $\delta$ quantization
and thus we do not see a direct contradiction. In particular, the bound on the Immirzi parameter
depends explicitly on the constant value of $\delta$ and so it is not clear if a bound exists
in the improved quantization. In addition, both papers evolve Gaussian semi-classical states
with the quantum difference equation which seem to imply that semi-classicality breaks down near the
singularity. The numerical results are based on using the triad component $p_c$ as an internal
clock, however, our results suggest that $p_c$ bounces and hence would not play the role
of a good clock. Thus to reconcile our results with those of \cite{Cartin:2006yv, Rosen:2006bga}, the numerical
evolutions would probably require different initial conditions to incorporate both the expanding and
contracting parts of the wave-packets at the initial instance in ``time''. This issue is solved
in the isotropic case by including a scalar field that plays the role of a global internal clock.
We thus feel that our results are not in conflict with \cite{Cartin:2006yv, Rosen:2006bga}, and
that all of our works requires further development of the quantum dynamics including the specification of
a good internal clock, physical inner product, and Dirac observables.

Additionally, the validity of the Kantowski-Sachs description of the interior
could be questioned. Boundary conditions must be provided for the numerical evolution
at the horizon which are prescribed using classical solutions. However, this could be
modified when considering a true quantization of the full spacetime. In particular, the
improved quantization scheme described in section~\ref{gooddelta} leads to
modifications at the horizon which may require input from the full quantization.
Additionally, by considering the homogeneous interior, we can not probe
a realistic collapse scenario. 

Therefore, an interesting next step would be to apply the techniques described
here to the inhomogeneous spherically symmetric quantization. The results
described in~\cite{Campiglia:2007pr} provide the initial steps at an
amenable quantization, though the results presented there are only valid in
the exterior. If the results can be extended to cover the whole spacetime,
then an effective dynamics analysis could be extended. With this model,
one could analyze a genuine collapse scenario.

\acknowledgments We thank Abhay Ashtekar and Roy Maartens for valuable discussion.
The work of CGB was supported by research grant BO 2530/1-1 of
the German Research Foundation (DFG). KV is supported by the Marie
Curie Incoming International Grant M1F1-CT-2006-022239.

\appendix

\section{Alternative Quantum Hamiltonian} \label{Alt}
Here we consider an alternate behavior for the $\delta$ factors that are more
heuristically motivated. In the language of~\cite{Bojowald:2007ra} this
corresponds to a lattice refinement model whereby the number of vertices is
proportional to the transverse area. In addition this scheme has been considered
in anisotropic Bianchi I models in~\cite{Chiou:2006qq,BianchiIeffective}.
A stability analysis  indicates that this scheme leads
to an unstable difference equation~\cite{Bojowald:2007ra} thus it may not represent
a good quantization scheme. An advantage of this scheme is
that the quantum difference equation is much simpler than in the previous one and
more amenable to a proper analysis of the true quantum dynamics although the
instability of the difference equation could spoil this.

In this scheme we have the $\delta$ factors behaving as
\ba
	\delta_b &=& \frac{\sqrt{\Delta}}{\sqrt{p_b}}\\
	\delta_c &=& \frac{\sqrt{\Delta}}{\sqrt{p_c}} \,.
\ea
Again the effective Hamiltonian is given by
\begin{align}
      \Ham_{\delta} = -\frac{N}{2 G \gamma^2} \Big[
      2 \frac{\sin  b \delta_b}{\delta_b} \frac{\sin \delta_c
c}{\delta_c} \sqrt{p_c} +
      \big(\frac{\sin^2\negmedspace b \delta_b}{\delta_b^2} + \gamma^2\big)
      \frac{p_b}{\sqrt{p_c}} \Big],
      \label{d1a}
\end{align}
leading to  equations of motion  with lapse $N=\gamma \sqrt{p_c} \delta_b/(\sin b \delta_b)$
\ba
	\dot{c} &=& c \,\cos c \delta_c - 3 \,\frac{\sin c \delta_c}{\delta_c} \\
	\dot{p}_c &=& 2 p_c \cos c \delta_c \\
	\dot{b} &=& - \frac{3 \sin b \delta_b}{4 \delta_b}+ \frac{b \cos b \delta_b}{4}
	-\frac{\gamma^2 \delta_b}{4 \sin b \delta_b	}
	-\frac{\gamma^2 \,b\, \delta_b^2  \cos b \delta_b}{4\sin^2 b \delta_b} \\
	\dot{p}_b &=&  \frac{1}{2} p_b \cos b \delta_b - \frac{\gamma^2 \,p_b \,\delta_b^2 \cos b \delta_b}{2 \sin^2 b \delta_b}
\ea

Again we can examine where the quantum corrections are appreciable by determining
where $b \delta_b$ and $c \delta_c$ are on the order of one. From the classical
behavior, both $b \delta_b$ and $c \delta_c$ blow up at the singularity and thus quantum effects
are expected there. At the horizon, $b \delta_b$ vanishes and $c \delta_c$ is small
provided $\Delta$ is small. Hence with this quantization scheme, the effects near the horizon are
minimal. The resulting numerical solutions are qualitatively very similar to those
obtained for the constant $\delta$ quantization scheme. As discussed above, the
parameter $p_b^{(0)}$ controls the mass of the second black hole in the effective
quantum theory which can be tuned to give a symmetric spacetime.
We note that similarly to the other quantization schemes, the singularity is resolved dynamically.

\begin{figure}[!ht]
\noindent
\begin{minipage}[h]{.48\linewidth}
\centering\epsfig{figure=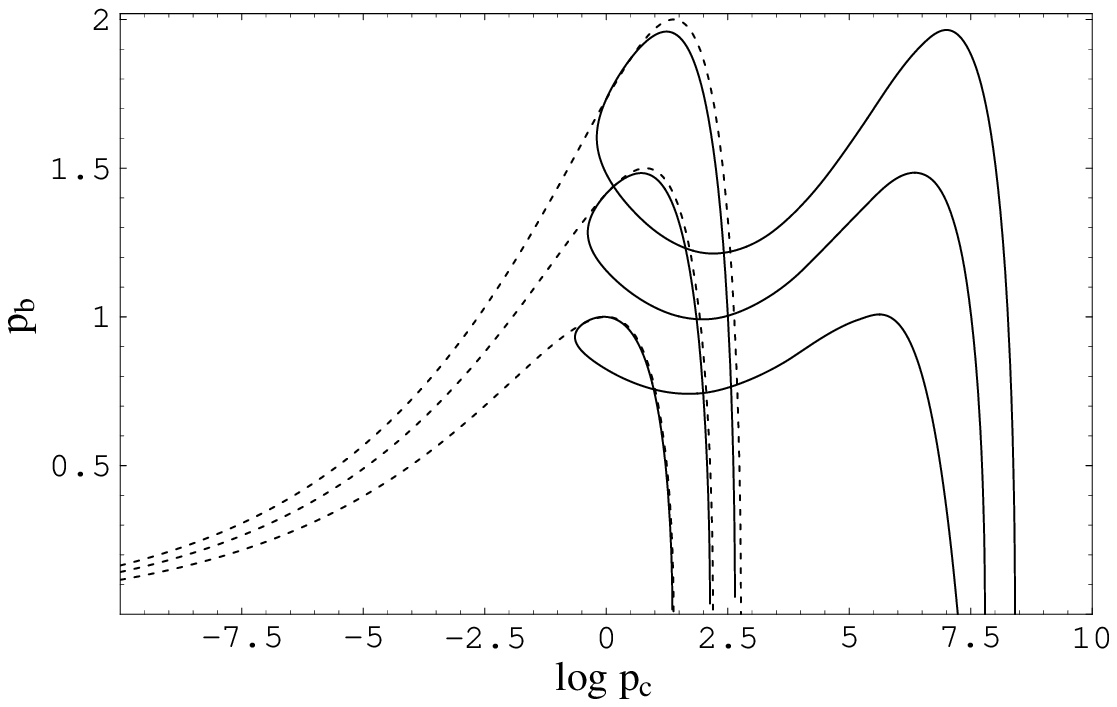,width=\linewidth}
\end{minipage}\hfill
\begin{minipage}[h]{.48\linewidth}
\centering\epsfig{figure=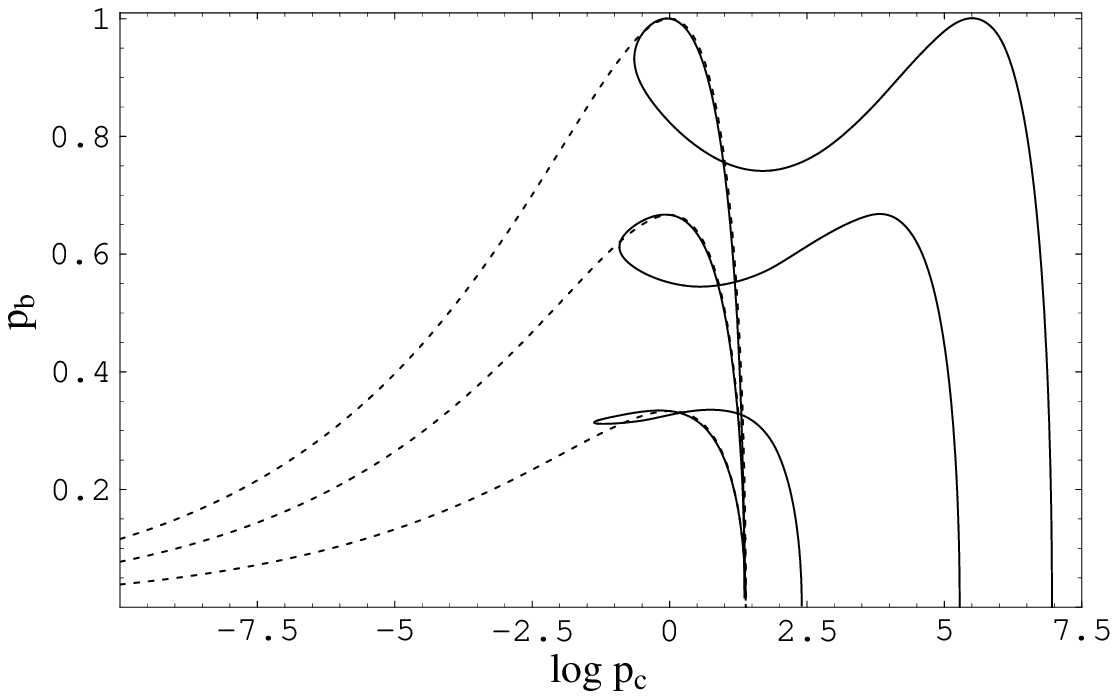,width=\linewidth}
\end{minipage}
\caption{These figures show the effective plots of $p_b = p_b (p_c)$. The dashed lines
represent the classical solution while the solid lines correspond to the solution of
the effective quantum theory. The initial singularity is removed and the spacetime is
connected to
 a second black hole.  The three plots on the left show the solution for varying
mass parameter, $m = 1; 1.5; 2$ respectively. On the right $p_b^{(0)}$ is varied in the
three plots, $p_b^{(0)} = 1; 2/3; 1/3$ respectively.}
\label{fig:deltabad1}
\end{figure}

\end{document}